# An Epistemic View of Quantum Communication

Subhash Kak
Curtin University, Perth and OSU, Stillwater

ABSTRACT
This paper presents an epistemological perspective on quantum communication between parties that highlights the choices that must be made in order to send and obtain information. The notion of information obtained in such a communication is a property associated with the observers and while dependent on the nature of the physical system its fundamental basis is epistemic. We argue that the observation process is in accord with the principle of psychophysical parallelism that was used by Bohr, von Neumann, Schrödinger and others to establish the philosophical basis of complementarity but has since fallen out of fashion. This principle gave coherence to the original Copenhagen Interpretation without which the latter has come to have some sort of an *ad hoc* character.

1. INTRODUCTION

A formal theory comes with its well-defined entities and rules of analysis. In addition, certain other entities are usually implicit in the theory but are not formally defined. These implicit entities and the underlying assumptions about reality, together with the different ways abstract entities may be mapped to intuitive notions, leads to divergent interpretations of the theory. Such divergences are particularly true for quantum theory for which the interpretations include the Copenhagen, stochastic evolution, consistent histories, transactional, QBism, MWI and so on (e.g. [1]-[6]). Roughly these interpretations fall into the epistemic and the ontic views where in the epistemic one is speaking of the knowledge obtained from the experiment without going into the ultimate nature of reality and in the ontic one is describing reality as a particular assemblage of objects. These views are so far from each other that their synthesis is impossible.

Even though the epistemic Copenhagen Interpretation has been the dominant view amongst quantum theorists [6], it has been criticized for being at variance with the contemporary program of science of finding the ontic (ontological) basis of reality. This basis is sought in the structure of being (in its physical embodiment) and that of its becoming (e.g. [7]) as well as other algorithmic models of physics and intelligence. Physicists attempt to describe a physical system mathematically and logically in a manner that can describe its evolution and in such an ontological description there is no place for observers. This program has grown hand in hand with the deepening use of computers in society and there are models where the unfolding of the universe itself is seen in terms of the workings of a computer program.

The ontic understanding of reality becomes problematic when one brings in information into the mix, as has been done rather extensively in modern physics. This is because information implies the existence of a mind, which is outside of the realm of physics. The study of mind is normally done using the tools and concepts of psychology and neuroscience. Standard neuroscience accepts the doctrine of an identity of brain and mind. In this view, mind emerges from the complexity of the interconnections and its behavior must be completely described by the corresponding brain function leaving no room for agency of the individual [8]. No specific neural correlate of consciousness has been found [9], [10]. There are also attempts to ascribe certain counterintuitive characteristics of the mind to underlying quantum processing [11], [12] but as it still implies a machine paradigm it cannot be the complete explanation even if quantum mechanics is shown to play a role in brain processes.

The perspective of epistemology presents a way to highlight the differences in the implicit assumptions. As the study of the nature of knowledge, epistemology is of relevance in examining interpretations of theory and the case of quantum information in a communications setting makes its conceptual basis most clear. We agree with the philosopher Fred Dretske who argued that [13] "A more precise account of information will yield a more creditable theory of knowledge. Maybe … communication engineers can help philosophers with questions raised by Descartes and Kant."

Information in a communication involves two things: first, commonalities in the vocabulary of communication between the two parties; and second, the capacity to make choices. The commonality of vocabulary requires that the underlying abstract signs used by the parties be shared which stresses the social aspects of communication. The capacity to make choice means agency that has no place in a world governed by closed laws unless one considers psychophysical parallelism that excludes causal interaction between mind and body. In the view of such parallelism, mental and physical phenomena are two aspects of the same reality like two sides of a coin.

Note that psychophysical parallelism was a dominant philosophical view in Europe in the late nineteenth and early twentieth centuries but now has been relegated to the margins [14]. According to Moritz Schlick, who was the leader of the Vienna Circle of Logical Positivists in the 1930s, psychophysical parallelism is the "epistemological parallelism between psychological conceptual system on the one hand and a physical conceptual system on the other. The 'physical world' *is* just the world that is designated by means of the system of quantitative concepts of the natural sciences." [15] The idea is in an old one having been first enunciated as *samavāya* (inherence) in the Vaiśeṣika Sūtra of Kaṇāda in India [16] and later in Europe by Leibniz.

Now consider how choices are made and how these choices are intelligible. With Heidegger, one may speak of the difference between the ontical and the ontological where the first is concerned with facts about objects and the second is concerned with the meaning of Being, with how objects are intelligible as entities. According to Heidegger [17], "Basically, all ontology, no matter how rich and firmly compacted a system of categories it has at its disposal, remains blind and perverted from its ownmost aim, if it has not first adequately clarified the meaning of Being, and conceived this clarification as its fundamental task."

There are usually several unstated assumptions regarding the process of obtaining information from an experimental situation that involve the nature of the observer. Specifically, we endow the observer with the capacity to make intelligent classifications and choices, either directly or through the agency of instruments and computing devices, which are not a part of the formal framework that describes the physical processes being investigated. It is interesting that some interpretations strive to take out the observer from the framework, without explaining how the central role of the selectivity in the observation process is to be explained.

In this paper we first review the problem of observation in epistemic and ontic interpretations of quantum theory presenting the key points of Bohr, von Neumann, and Schrödinger that views the epistemic understanding emerging from the principle of psychophysical parallelism. Next we examine the question of information in classical and quantum settings highlighting how its definition in the framework of ensembles requires an epistemic basis. We argue that the Copenhagen Interpretation provides the best resolution to the problems associated with information.

2. THE OBSERVATION PROCESS AND COMPLEMENTARITY

We first consider the epistemic Copenhagen Interpretation in which the physical universe is separated into two parts, the first part is the system being observed, and the second part is the human observing agent, together with the instruments. The agent is therefore an extended entity that described in mental terms and it includes not only his apparatus but also instructions to colleagues on how to set up the instruments and report on their observations. The Heisenberg cut (also called the von Neumann cut) is the hypothetical interface between quantum events and the observer's information, knowledge, or awareness. Below the cut everything is governed by the wave function, whereas above the cut one must use classical description.

Although the arbitrariness of the cut has come in for criticism and spurred the development of other interpretations, it is a device for aggregating the effects of the mind or minds associated with the observational regime and it is a reasonable way to separate the inanimate from the animate especially since the brain itself may be viewed as a machine. Bohr stressed the elusive separation between subject and object [18]:

> The epistemological problem under discussion may be characterized briefly as follows: For describing our mental activity, we require, on one hand, an objectively given content to be placed in opposition to a perceiving subject, while, on the other hand, as is already implied in such an assertion, no sharp separation between object and subject can be maintained, since the perceiving subject also belongs to our mental content.

Von Neumann describes the principle thus [6]: "[I]t must be possible so to describe the extra-physical process of the subjective perception as if it were in reality in the physical world -- i.e., to assign to its parts equivalent physical processes in the objective environment, in ordinary space." He adds further:

> The boundary between the two is arbitrary to a very large extent. In particular we saw in the four different possibilities in the example above, that the observer in this sense needs not to become identified with the body of the actual observer: In one instance in the above example, we included even the thermometer in it, while in another instance, even the eyes and optic nerve tract were not included. That this boundary can be pushed arbitrarily deeply into the interior of the body of the actual observer is the content of the principle of the psycho-physical parallelism -- but this does not change the fact that in each method of description the boundary must be put somewhere, if the method is not to proceed vacuously, i.e., if a comparison with experiment is to be possible. Indeed experience only makes statements of this type: an observer has made a certain (subjective) observation; and never any like this: a physical quantity has a certain value.

> Now quantum mechanics describes the events which occur in the observed portions of the world, so long as they do not interact with the observing portion, with the aid of the process 2, but as soon as such an interaction occurs, i.e., a measurement, it requires the application of process 1. The dual form is therefore justified. However, the danger lies in the fact that the principle of the psycho-physical parallelism is violated, so long as it is not shown that the boundary between the observed system and the observer can be displaced arbitrarily in the sense given above.

The above quotes make it clear that psychophysical parallelism is not equivalent to brain-mind identity of neuroscience in which the mind is an emergent property with neural structures as ground thus admitting a causal link going from biology to the mind.

The question of interaction between mental states and the wave function was addressed in the Copenhagen Interpretation (CI) [2] in which the wave function is properly understood epistemologically, that is, it represents the experimenter's knowledge of the system, and upon observation there is a change in this knowledge. Operationally, it is a dualist position, where there is a fundamental split between observers and objects. The placement of the cut between the subject and the object is arbitrary to the extent it depends on the nature of the interaction between the two.

In the ontic view of the wave function as in the Many Worlds Interpretation (MWI), there is no collapse of the wave function and the interaction is seen through the lens of decoherence, which occurs when states interact with the environment producing entanglement [19]. By the process of decoherence the system makes transition from a pure state to a mixture of states that observers end up measuring. The problem of collapse of the wave function is sidestepped by speaking of interaction between different subsystems. But since the entire universe is also a quantum system, the question of how this whole system splits into independent subsystems arises. It would seem that the splitting into subsystems is itself an observational choice, rather than fundamental. This splitting serves about the same function as the Heisenberg cut of CI. Furthermore, such an ontic view has no place for minds, which can at best be taken as traces of mathematical operations thus ruling out agency.

Finally, the principle of psychophysical parallelism is consistent with complementarity and indeed the inspiration for it [20]. Bohr argued that the consideration of the biological counterpart to the observation of the relation between mind and body does not become part of an infinite regress. He argued that [21] "We have no possibility through physical observation of finding out what in brain processes corresponds to conscious experience. An analogy to this is the information we can obtain concerning the structure of cells and the effects this structure has on the way organic life displays itself.… What is complementary is not the idea of a mind and a body but *that* part of the contents of the mind which deals with the ideas of physics and the organisms and *that* situation where we bring in the thought about the observing subject."

Schrödinger implicitly invoked the principle in describing the state function of a quantum state (psi) as representing our knowledge about the system. He said [22]:

> Reality resists imitation through a model… We have nothing but our reckoning scheme, i.e., what is a *best possible* knowledge of the object. The psi-function … is now the means for predicting probability of measurement results. In it is embodied the momentarily-attained sum of theoretically based future expectation, somewhat as laid down in a *catalog*… [This] the catalog of expectations is initially compiled. From then on it changes with time, just as the state of the model of classical theory, in constrained and unique fashion… For each measurement one is required to ascribe to the psi-function (= the prediction-catalog) a characteristic, quite sudden change, which *depends on the measurement result obtained*, and so *cannot be foreseen*; from which alone it is already quite clear that this second kind of change of the psi-function has nothing whatever in common with its orderly development *between* two measurements… And indeed because one might never dare impute abrupt unforeseen changes to a physical thing or to a model, but because in the realism point of view observation is a natural process like any other and cannot *per se* bring about an interruption of the orderly flow of natural events.

It is clear that Schrödinger is stressing the epistemic nature of the state function. Elsewhere, he presents the psychophysical parallel basis of this claim in a clearer form: "Consciousness cannot be accounted for in physical terms. For consciousness is absolutely fundamental. It cannot be accounted for in terms of anything else." [23]

The complementarity of aspects, such as wave and particle, is a consequence of the kind of measurement that is made which emanates from the choice made by the observer and it is not the description of an ontologically defined entity in two equivalent forms as in the representation of a number directly or in terms of its inverse sequence [24]. The experimenter is not describing reality ontologically; rather, he is obtaining knowledge about it that is related to the nature of his interaction with the system. The particle view is the one imposed on reality by the mind governed by a classical mode [25]. Nonlocality is an issue only if one takes the particle picture, together with local interaction, to be underlying reality and, therefore, the violation of Bell's theorem by experiments does not imply a fundamental difficulty [26].

3. THE COMMUNICATIONS SETTING

We consider the problem of exchange of information between two parties. Figure 1 describes the communications context for the consideration of information [27]. It consists of a sender and a receiver together with an ensemble of signals (which could be letters). The statistical characteristics of the signals are known both to the sender and the receiver. There could be further relationship between the symbols and physical or abstract objects in which case one can also speak of a semantic content communicated through the transmissions.

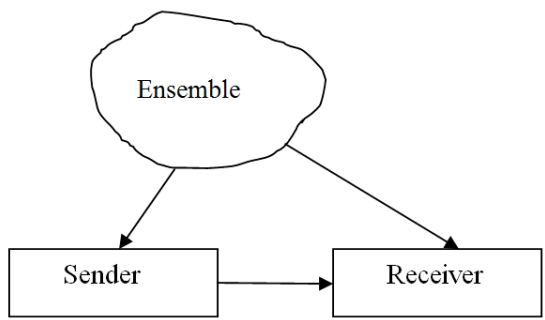

*Figure 1*. Exchange of information

Let the probabilities of the signals that are transmitted and then received by the sender, S, and the receiver, R, be $p(x_i)$ and $p(y_j)$, respectively (the discrete index refers to the specific signal being considered out of a list that ranges from 1 onwards).

These are the *a priori* probabilities associated with the sender, the receiver, and their world. The sender now chooses a specific one out of the ensemble and sends it to the receiver and repeats this process. The information exchanged between S and R is:

$$I(x_i, y_j) = -\log \frac{p(x_i | y_j)}{p(x_i)} \qquad (1)$$

and this information is always positive, if not zero. The informational entropy is, therefore:

$$I(X;Y) = \sum_{x,y} p(x,y) \log \frac{p(x,y)}{p(x)p(y)} \qquad (2)$$

Also,
$$I(X;Y) = H(X) - H(X|Y) \qquad (3)$$

We ask: what is the connection between probabilistic information provided by the entropy expressions above and the knowledge obtained by the receiving party? At the most basic level, the following claims may be made regarding the communication process:

1. There exist associations of data, which requires separating it from other data, and abstractions (input X and output Y) which are assumed without explaining how this is achieved.
2. Training set with correct classifications (or data typical of the ensemble) that involves different modes of behavior.
3. Classification task (as in AI and neural networks) with hierarchical levels of understanding.
4. Duality between the process of the identification of the ensemble (learning) and that of subsequent measurements.

Implicit in the identification of the ensemble is the mind and it is also implicitly acknowledged in the problem of classical information. The sharing of the ensemble must be part of a social process.

Now consider the communications context for the quantum case. Assume the sender and the receiver both are informed of the ensemble of states {ρ$_1$, ρ$_2$, ..., ρ$_n$} with probability {p$_1$, p$_2$, ..., p$_n$}. Every density operator may be viewed as a mixture of pure states

$$\rho = \sum_i \lambda_i |\varphi_i\rangle\langle\varphi_i| \qquad (4)$$

where $\lambda_i$ are the eigenvalues and $|\varphi_i\rangle$ are the eigenvectors of $\rho$. The entropy may be written as [12]:

$$S_n(\rho) = -\sum_i \lambda_i \log \lambda_i \qquad (5)$$

where $\lambda_i$ are the eigenvalues of $\rho$. Thus the measurements along the reference bases may be associated with probability values $\lambda_i$ in analogy with the classical entropy expression of $-\sum_i p_i \log p_i$, where the *i*th outcome, out of a given set, has probability $p_i$.

Operationally, classical and quantum information work differently. Given an unknown state distributed over two systems, consider how much information needs to be sent to transfer the full state to a system. In the classical case, partial information must always be positive, but in the quantum case it can be negative. If the partial information is positive, its sender needs to communicate this number of quantum bits to the receiver; if it is negative, then sender and receiver instead gain the corresponding potential for future quantum communication [28]. The idea of negative partial information makes sense only in the context of observers that are epistemologically connected.

Now consider the case of Figure 2 in which the sender has access to states in an ensemble and he is sending these to the receiver. We are interested in considering the entropy of this situation. If the sender is Nature, the receiver is the experimenter who is determining both the ensemble as well as the information that is associated with it.

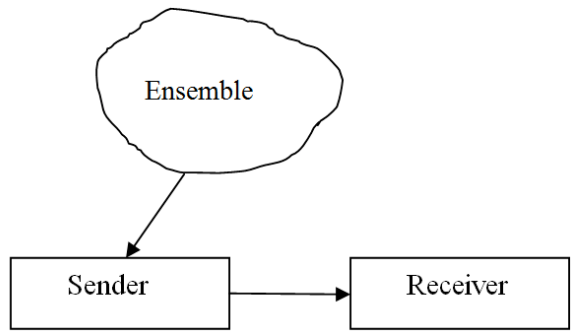

*Figure 2*. Directed information transfer

If the ensemble consists of a single unknown letter, the Shannon entropy associated with the ensemble is zero. The information in any single communication is also zero. Starting with the hypothesis that there are two potential states, the receiver will find in the nth test that the probability that it is a single letter is $1-2^{-n}$.

In contrast, in the quantum case, let the ensemble consist of a single pure state. We assume that the reference bases of the sending and the receiving parties are aligned without going into the question of the cost associated with that process. The von Neumann entropy associated with this case is zero. Nevertheless, the amount of information that the receiver can obtain is infinite [12]. This infinite information will be in terms of the specific phase information associated with the state which, theroretically, has infinite precision.

Parenthetically, an application of this is when two parties wish to share a random number. The sender codes the random number into the polarization angle of the many copies of the photons. The determination of the angle by the receiver will eventually transfer the random sequence to the receiver [12],[29].

The ability to obtain information from the unknown state implies a corresponding potential energy. The equivalence between energy and information is given by $kT\ln2$ (or about $0.69\ kT$) which is both the minimum amount of work needed to store one bit of binary information and the maximum that is liberated when this bit is erased, where $k$ is Boltzmann's constant and $T$ is the temperature of the storage medium in degrees Kelvin [30].

Since the unknown state may be assumed to be localized at a physical point, one can thereby conclude that energy is associated with space. This is restating the concept of zero-point energy that is normally derived using Heisenberg's Uncertainty Principle. The ability of obtain information from space would depend on how the experimenter interacts with it.

The consideration of information also requires choices that go into the formation of the ensemble that is used by the communicating parties. This is only possible by the observer making choices. The manner in which these choices are made will change the value of the entropy associated with the process of information exchange [12]. The choices establish that the information is epistemic. Without consideration of this aspect of information basis, we are confronted by difficulties such as the information paradox of cosmology [31].

4. DISCUSSION

The notion of psychophysical parallelism rules out the need of hidden-variable theories. According to it, quantum mechanics is an epistemic theory in which there is no need to introduce additional variables that will convert it into an ontic theory. The lack of experimental support for hidden-variable theories is to be expected within the framework of this parallelism. It is also not surprising that extensions to quantum theory cannot give more information about the outcomes of future measurements than quantum theory itself [32].

One must also assume that the psychological part of the psychophysical parallelism notion implies that there exists no specific correlate of consciousness in the brain (as it cannot have a physical basis). The quantum Zeno effect [33] does provide a mechanism on how observation can influence dynamics but it does not explain the ontological position of the observer.

5. REFERENCES
1. W. Heisenberg, Physics and Philosophy: the Revolution in Modern Science. George Allen & Unwin, London (1971)
2. J. von Neumann, The Mathematical Foundations of Quantum Mechanics. Princeton: Princeton University Press (1955)
3. W.M. Dickson, Quantum Chance and Nonlocality. Cambridge University Press (1998)
4. N. Harrigan and R.W. Spekkens, Einstein, incompleteness, and the epistemic view of quantum states. Found. Phys. 40, 125–157 (2010)
5. R.E. Kastner, The Transactional Interpretation of Quantum Mechanics. Cambridge University Press (2012)
6. M. Jammer, The Philosophy of Quantum Mechanics. New York: John Wiley (1974)
7. D. Bohm. A suggested interpretation of the quantum theory in terms of hidden variables. I. Phys. Rev. 85: 166-179 (1952)
8. R. Melzack, Phantom limbs, the self and the brain. Canadian Psychology 30, 1-16 (1989)
9. S. Zeki, The disunity of consciousness. Trends Cogn Sci 7, 214-218 (2003)
10. A. Kak et al., A three-layered model for consciousness states. NeuroQuantology 14, 166-174 (2016)
11. S. Kak, The three languages of the brain: quantum, reorganizational, and associative. In Learning as Self-Organization, K. Pribram and J. King (editors). Lawrence Erlbaum Associates, Mahwah, NJ, 185-219 (1996)
12. S. Kak, Quantum information and entropy. Int. Journal of Theoretical Physics 46, 860-876 (2007); S. Kak, State ensembles and quantum entropy. International Journal of Theoretical Physics 55, 3016-3026 (2016)
13. F. Dretske, Epistemology and information. In Handbook of the Philosophy of Science. Volume 8: Philosophy of Information, ed. P. Adriaans and J. van Benthem. Amsterdam: Elsevier (2008)
14. D. Ludwig, A Pluralist Theory of Mind. Springer (2015)

15. M. Schlick, General Theory of Knowledge. Springer(1918/1974)
16. S. Kak, The Nature of Physical Reality. Mt. Meru, Canada (2016)
17. M. Heidegger, Being and Time. Oxford: Basil Blackwell (1962)
18. N. Bohr, Wirkungsquantum und Naturbeschreibung. Die Naturwissenschaften 17, 483–486 (1929)
19. S. Kak, Veiled nonlocality and quantum Darwinism. NeuroQuantology 13, 10 – 19 (2015)
20. N. Bohr, Atomic Physics and Human Knowledge. Wiley (1958)
21. N. Bohr, Complementarity Beyond Physics. Elsevier (2013)
22. E. Schrödinger, Die gegenwärtige Situation in der Quantenmechanik. Naturwissenschaften 23, 823-828, (1935)
23. W. Moore, Schrödinger: Life and Thought. Cambridge University Press (1994)
24. S. Kak and A. Chatterjee, On decimal sequences. IEEE Trans. on Information Theory IT-27: 647 – 652 (1981)
25. S. Kak, From the no-signaling theorem to veiled nonlocality. NeuroQuantology 12, 12-20 (2014)
26. E. Joos et al., Decoherence and the Appearance of a Classical World in Quantum Theory. Springer (2003)
27. C.E. Shannon, A mathematical theory of communication. Bell Syst. Tech. J. 27, 379–423, 623–656 (1948)
28. M. Horodecki, J. Oppenheim, A. Winter, Partial quantum information. Nature 436, 673-676 (2005)
29. S. Kak, A three-stage quantum cryptography protocol. Foundations of Physics Letters 19, 293-296 (2006)
30. R. Landauer, The physical nature of information. Phys. Lett. A 217,188-193 (1996)
31. R. Penrose, The Road to Reality. Vintage Books (2004)
32. R. Colbeck and R. Renner, No extension of quantum theory can have improved predictive power. Nature Communications 2 (8) (2011)
33. B. Misra and E.C.G. Sudarshan, The Zeno's paradox in quantum theory. Journal of Mathematical Physics 18, 756–763 (1977)